# Dependence of Levitation Force on Frequency of an Oscillating Magnetic Levitation Field in a Bulk YBCO Superconductor

Hamilton Carter[a], Stephen Pate[b], George Goedecke[c]

a. Department of Physics, New Mexico State University, Las Cruces, New Mexico 88003, United States, hcarter3@nmsu.edu

b. Department of Physics, New Mexico State University, Las Cruces, New Mexico 88003, United States, pate@nmsu.edu

c. Department of Physics, New Mexico State University, Las Cruces, New Mexico 88003, United States, ggoedeck@nmsu.edu

The dependence of the magnetic field strength required for levitation of a melt textured, single domain YBCO superconductor disc on the frequency of the current generating the levitating magnetic field has been investigated. The magnetic field strength is found to be independent of frequency between 10 and 300 Hz. This required field strength is found to be in good experimental and theoretical[1] agreement with the field strength required to levitate the same superconductor with a non-oscillating magnetic field. Hysteretic losses within the superconductor predicted by Bean's critical-state model[2] were also calculated. The measured data rules out any significant Bean's model effects on the required levitation field strength within the measured frequency range.

## 1. Introduction

Levitation systems utilizing permanent magnets, or DC electromagnets and YBCO bulk superconductors have been well documented and characterized[1][3][4]. Bulk YBCO levitation systems utilizing oscillating magnetic fields have been reported[5][6], but not characterized in detail. In this experiment the minimum magnetic field strength required for levitation of a bulk YBCO disc was measured with respect to the frequency of the field. Data was obtained for frequencies between 10 and 300 Hz.

## 2. Material and Methods

The YBCO superconductor used for the experiment is a disk type, single domain, melt-texture-growth bulk sample with a diameter of 14 mm, and a thickness of 6mm from CAN Superconductors. The sample has a critical temperature of 90 K and was zero-field cooled by being suspended in boiling liquid nitrogen at a temperature of 77 K.

The experimental procedure is as follows:

- The YBCO bulk is zero-field cooled for three minutes by being suspended in a liquid nitrogen reservoir directly above the electromagnet used to provide the oscillating magnetic field.
- The frequency of the alternating current supplied to the electromagnet is measured and recorded using an oscilloscope.
- The driving current is supplied to the electromagnet and slowly ramped upwards beginning at a peak value of zero volts. When levitation of the superconductor is first detected, the peak value of the sinusoidal voltage waveform created by a three turn pick-up coil wound directly around the top of the electromagnet is measured and recorded. This value is used to calculate, (as described below), the root mean square magnitude of the magnetic field required for levitation.

The experimental apparatus is shown in Figure 1. The superconductor is supported by a plastic boom hinged on a stretched steel 0.055 inch diameter piano wire. This portion of the apparatus is based on the hysteretic levitation measuring device reported by Weeks[7]. The oscillating magnetic levitation field is provided by an electromagnet situated directly below the liquid nitrogen reservoir. The sinusoidal current for the electromagnet is supplied by a signal generator and audio frequency power amplifier.

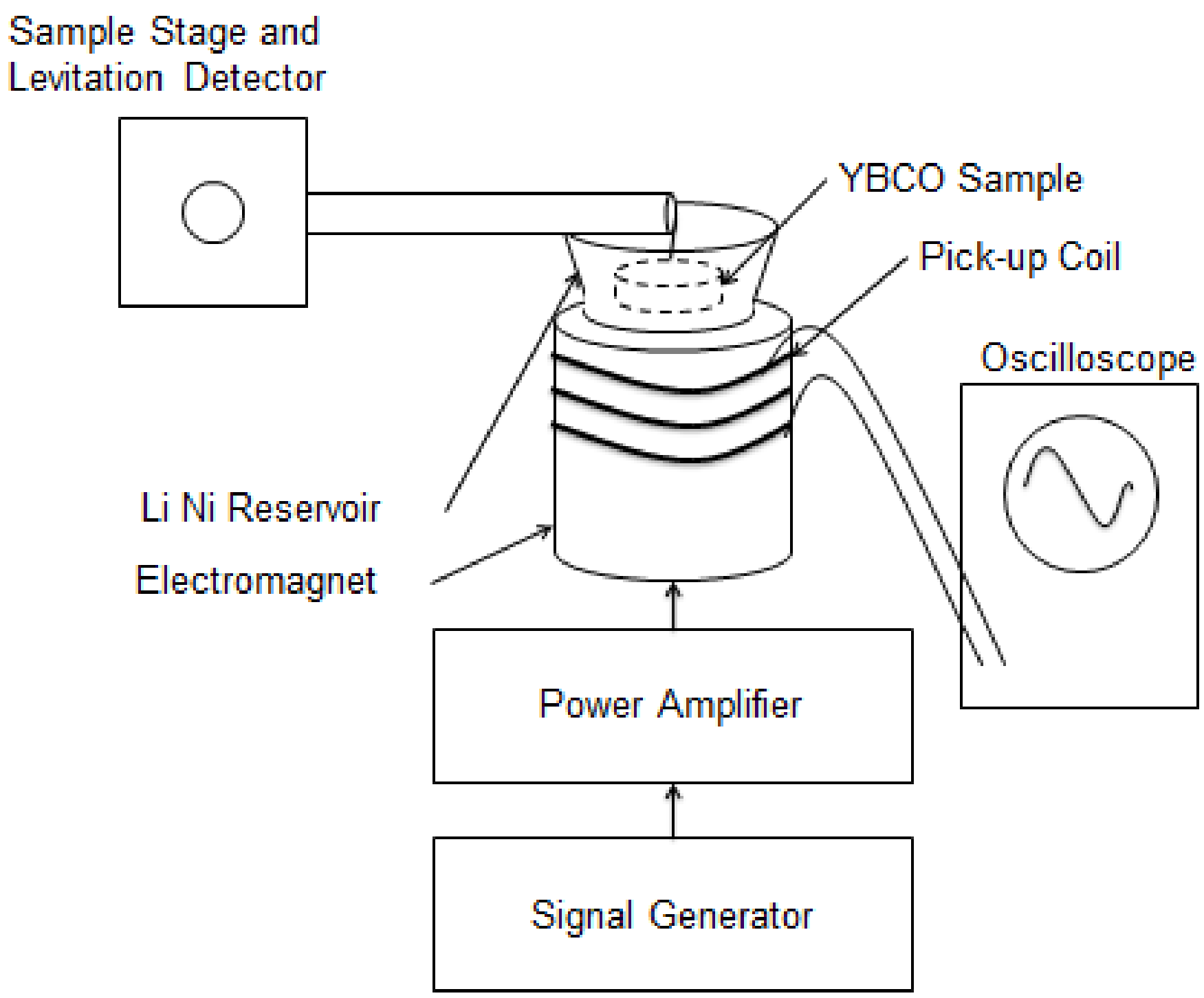

Figure 1

Levitation is detected by means of a small switch, consisting of aluminum foil at the base of the boom that makes a connection between two wires directly below the boom. These two wires and the edge of the liquid nitrogen reservoir support the boom. When the superconductor is levitated, the boom swings free of the two supporting wires breaking the electrical connection. The broken connection is detected by a microcontroller which activates a light emitting diode, (LED), indicating that levitation has been achieved.

The hinge, support boom, and levitation detector are all installed on top of a laboratory jack. For maximum detector sensitivity, with the sample end of support boom resting in the fixed liquid nitrogen reservoir, the jack is lowered until the levitation LED is activated indicating that the boom has just lifted above the switch. The jack is then slowly raised until the LED is extinguished. This calibrates the system so that it is sensitive to the minimum possible amount of lift provided by the levitation of the superconductor.

## 3. Theory/Calculation

A pick-up coil was used to indirectly measure the oscillating magnetic field via induction. Due to the oscillating magnetic field, this provided a simple, and more precise measurement than could be achieved with the available Hall probe. The peak value of the waveform measured from the pick-up coil was converted to a magnetic field using Faraday's law of induction:

$$\varepsilon = -\frac{d\Phi_B}{dt} \tag{1}$$

Where $\varepsilon$ is the voltage induced in the pick-up coil by the flux, $\Phi_B$, of the electromagnet. The measured voltage is sinusoidal with frequency $\omega$ and a peak value of $V_{\max}$ giving:

$$V_{\max}\sin(\omega t) = -\frac{d\Phi_B}{dt} \tag{2}$$

Integrating both sides, we arrive at:

$$\frac{V_{\max}}{\omega}\cos(\omega t) = \Phi_B + C \tag{3}$$

The constant of integration $C$ is set to zero because the pick-up coil voltage oscillates around zero Volts.

Equation 3 shows that the peak value of the magnetic flux is

$$\Phi_{BMax} = \frac{V_{\max}}{\omega} \quad . \tag{4}$$

Using the area of the pick-up coil, $A$, and the number of turns, $T$, the peak magnetic flux can be used to calculate the peak magnetic field as

$$B_{Max} = \frac{\Phi_{BMax}}{AT} \quad . \tag{5}$$

Teshima et al. [4] derived the maximum levitation force on an ideal superconductor due to a *static* magnetic field

$$F = \frac{AB^2}{2\mu_o} = mg\,, \qquad (6)$$

where A is the surface area of the superconductor, B is the magnetic field, m is the combined mass of the superconductor, sample holder, and support boom in our cases and g is the acceleration of gravity. To account for the weight born by the hinge, the entire unit was weighed in-situ using a balance beam scale attached to the sample holder while it rested at the same height as the edge of the liquid nitrogen reservoir. Under the reasonable assumption that the levitation force still depends on $B^2$ for oscillating fields, (in which case, $B^2$ would be the mean squared field), equation 6 reveals that the magnetic field strength required for levitation should be independent of frequency.

## 4. Results and Discussion

The measured peak pick-up coil voltages vs. the frequency of the levitating magnetic field are shown in figure 2.

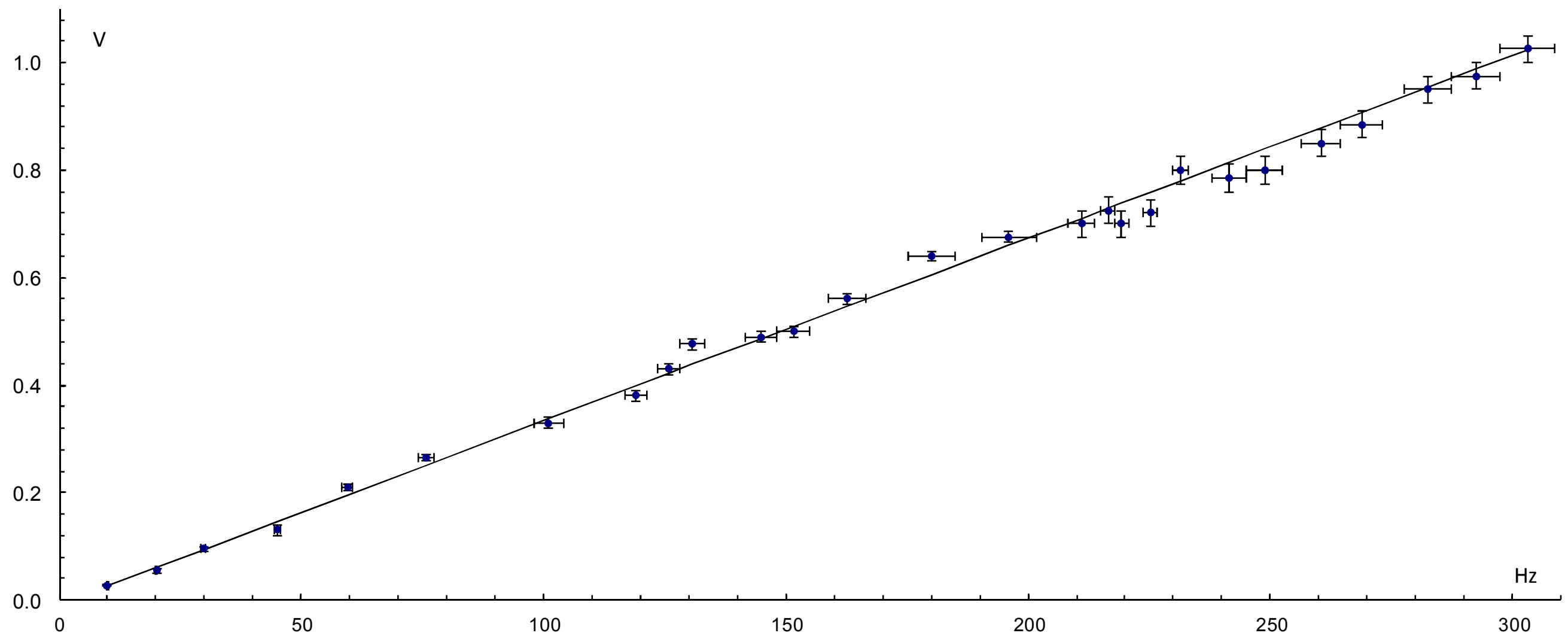

Figure 2

The data was taken using an analog oscilloscope. The error bars reflect a $\pm 0.1$ division uncertainty when reading both the time and voltage scales of the oscilloscope. The data shows a linear dependence between the pick-up coil voltage at levitation and the frequency of the current driving the electromagnet. A chi squared fit was performed on the data yielding, $V = 0.0034f - 0.01$, with $\chi^2 / d.o.f. = 1.56$.This indicates that within the measured frequency range, the value of the peak magnetic flux required for levitation remains constant and independent of frequency. Using equation 5, the slope of the chi squared fit line, $3.4mV / Hz \pm 0.03mV / Hz$ produces a root mean square magnetic field value of $372 \pm 3$ Gauss.

A magnetic field map of the solenoid pole piece and the overlaying superconductor, (dashed line), is shown in Figure 3.

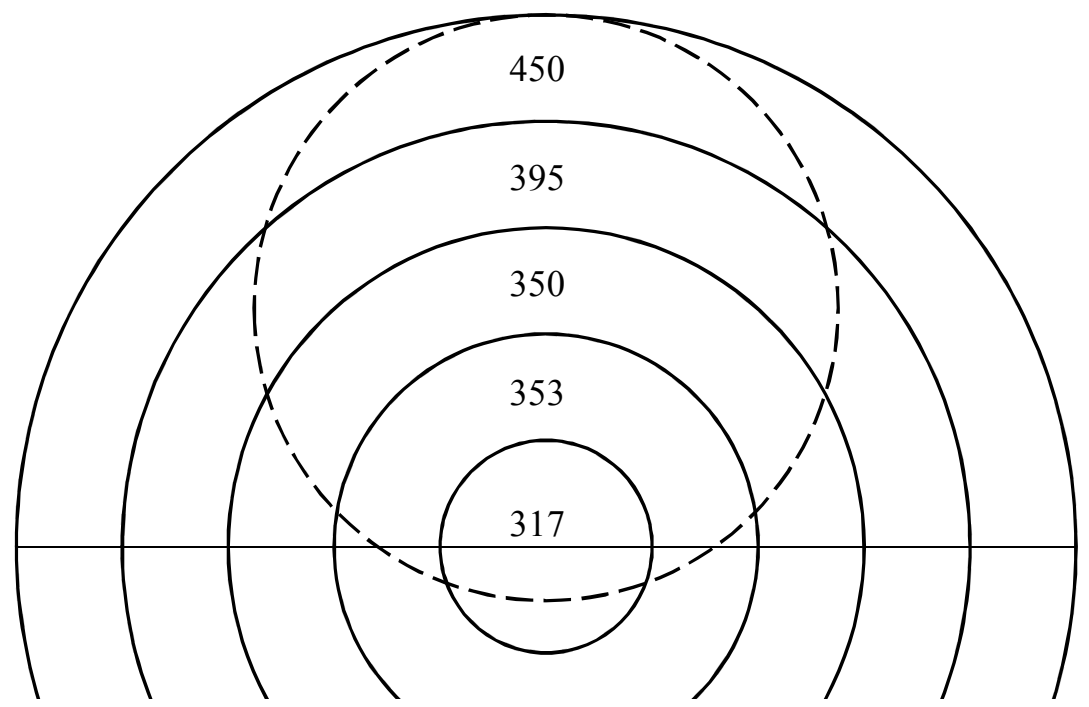

Figure 3

The mapped values were measured using a Hall probe and correspond to the magnetic field present at the onset of DC levitation of the superconductor. A field measurement was taken directly on the pole piece every 0.254 cm along the vertical axis of the superconductor. The values were recorded in Gauss with an uncertainty of $\pm 5$ Gauss. The complete field map was extrapolated by assuming that the measured values extended around the pole piece in concentric circles. The magnetic field beneath the area covered by the superconductor was calculated as a weighted average, based on the area of the intersection of each concentric circle with the superconductor, and gave a result of 377 Gauss. The value of $372 \pm 3$ Gauss obtained from the chi squared fit is in agreement with this weighted average.

Using the weight, 8.99 grams, of the superconductor, the sample holder, and support boom measured with a balance placed at the level of the liquid nitrogen reservoir we arrive at a required levitation force of 0.088 Newtons which falls within the maximum levitation force of 0.16 Newtons calculated using equation 6.

Bean's model[2] states that the hysteretic loss caused by an oscillating magnetic field that is much less than the superconductor's first critical field can be modeled as a surface loss and is given by

$$W_s = \frac{5H_o^{\,3}}{12\pi^2 J_c} ergs/cm^2/cycle, \qquad (7)$$

where $H_o$ is the magnitude of the oscillating levitation field, (511 Gauss), and $J_c$ is the superconductor's critical current density, (about 10,000 $A/cm^2$ for YBCO superconductors). Aravind et al.[8] report a value for the specific heat capacity of a YBCO superconductor at 77° K of $0.96\,J/cm^3/K$. We used equation 7 to determine the hysteretic loss per second shown in Table 1. Using the reported specific heat capacity adjusted for the volume of our superconductor, and making the worst case assumption that none of the heat from the superconductor is conducted to the liquid nitrogen bath, equation 7 yield the maximum temperature change values shown in Table 1 below.

| Frequency | Loss/second | Temperature Change/second |
|---|---|---|
| 10 Hz | 0.789 mJ | 0.88 mK/s |
| 300 Hz | 23.7 mJ | 26.7 mK/s |

Table 1

Using the rough estimate that our sample changes from the normal state at room temperature to superconducting after 2 minutes in a liquid nitrogen bath provides a cooling rate of -1.72 K/s. This rate is more than sufficient to conduct away heat created by hysteretic loss.

## 5. Conclusions

The magnetic field strength of an oscillating magnetic field required to levitate a YBCO bulk superconductor is shown to be in good agreement with the field strength required to levitate the same superconductor with a constant magnetic field. No dependence on the frequency of the levitating field was detected within the experimental error of our instrumentation. The hysteretic loss predicted by Bean's critical state model was calculated and has a negligible effect at the frequencies measured.

## 6. Acknowledgements

Funding for this research was provided by the New Mexico State University Space Grant Consortium.

Hamilton Carter was also funded by NASA's Space Grant College and Fellowship Program.

**Figure Captions**

Figure 1 Schematic diagram of the apparatus for the measurement of the oscillating magnetic field strength required for levitation of the superconducting YBCO sample.

Figure 2  Pick-up coil voltage vs. frequency
Figure 3 Solenoid pole piece magnetic field mapping

**Tables**

| Frequency | Loss/second | Temperature Change/second |
|---|---|---|
| 10 Hz | 0.789 mJ | 0.88 mK/s |
| 300 Hz | 23.7 mJ | 26.7 mK/s |

Table 1